\documentclass[Conference,a4paper]{IEEEtran}
\IEEEoverridecommandlockouts
\usepackage{color}
\usepackage{graphicx}
\usepackage{amsmath}
\usepackage{amssymb}
\usepackage{algorithm}
\usepackage{algorithmic}
\usepackage{amsmath}
\usepackage{multirow}
\usepackage{booktabs}
\usepackage{array}
\usepackage{amsthm}
\usepackage{stfloats}
\usepackage{caption}
\usepackage{subfigure}
\usepackage{bm}
\usepackage{booktabs}
\usepackage{setspace}

\newcommand{\be}{\begin{equation}}
\newcommand{\ee}{\end{equation}}
\newcommand{\bea}{\begin{eqnarray}}
\newcommand{\eea}{\end{eqnarray}}
\newcommand{\ba}{\begin{array}}
\newcommand{\ea}{\end{array}}

\title{Channel Estimation for Practical IRS-Assisted OFDM Systems
\thanks{$^{\ast}$ Corresponding author.}
\thanks{This work is supported in part by the National Natural Science Foundation of China (Grant No. 61971088, 62071083, U1808206, and U1908214), in part by the Fundamental Research Funds for the Central Universities (Grant No. DUT20GJ214 and DUT20RC(3)029), in part by Dalian Science and Technology Innovation Project (Grant No. 2020JJ25CY001),  and  in part by the Open Research Fund of National Mobile Communications Research Laboratory, Southeast University (Grant No. 2021D08).}
}
\author{\IEEEauthorblockN{Wanning Yang$^{\dag}$, Hongyu Li$^{\dag}$, Ming Li$^{\dag\ddag\ast}$, Yang Liu$^{\dag}$, and Qian Liu$^{\dag}$
\vspace{-0.0 cm} }\\
\IEEEauthorblockA{$^{\dag}$Dalian University of Technology, Dalian, Liaoning 116024, China
 \\ E-mail: \texttt{\{yangwanning, hongyuli\}@mail.dlut.edu.cn \\ \{mli, yangliu{\_}613, qianliu\}@dlut.edu.cn} } \\
\IEEEauthorblockA{$^{\ddag}$National Mobile Communications Research Laboratory  \\  Southeast University, Nanjing, Jiangsu, 210096, China}}

\begin{document}

\maketitle
\pagestyle{empty}
\thispagestyle{empty}

\begin{abstract}
Intelligent reflecting surface (IRS), composed of a large number of hardware-efficient passive elements, is deemed as a potential technique for future wireless communications since it can adaptively enhance the propagation environment. In order to effectively utilize IRS to achieve promising beamforming gains, the problem of channel state information (CSI) acquisition needs to be carefully considered. However, most recent works assume to employ an ideal IRS, i.e., each reflecting element has constant amplitude, variable phase shifts, as well as the same response for the signals with different frequencies, which will cause severe estimation error due to the mismatch between the ideal IRS and the practical one. In this paper, we study channel estimation in practical IRS-aided orthogonal frequency division multiplexing (OFDM) systems with discrete phase shifts. Different from the prior works which assume that IRS has an ideal reflection model, we perform channel estimation by considering amplitude-phase shift-frequency relationship for the response of practical IRS. Aiming at minimizing normalized-mean-square-error (NMSE) of the estimated channel, a novel IRS time-varying reflection pattern is designed by leveraging the alternating optimization (AO) algorithm for the case of using low-resolution phase shifters. Moreover, for the high-resolution IRS cases, we provide another practical reflection pattern scheme to further reduce the complexity. Simulation results demonstrate the necessity of considering practical IRS model for channel estimation and the effectiveness of our proposed channel estimation methods.
\end{abstract}


\maketitle

\section{Introduction}
Intelligent reflecting surface (IRS), as a promising technology to enhance propagation in wireless communication systems, has recently attracted significant attention. The IRS is generally composed of a great number of energy-efficient passive elements \cite{01}, e.g., phase shifters, which can dynamically modify the phase or/and amplitude of the incident signal and effectively improve propagation environment without additional hardware cost. Appealed by the advantage of IRS, corresponding applications for a wide range of communication scenarios, e.g. coverage enhancement \cite{001}, data transmission rate improvement \cite{ao},  and secure communication \cite{secure}, have been intensively investigated.

To completely achieve the appealing advantages brought by the IRS, the issue of accurate channel state information (CSI) acquisition needs to be deliberately considered. However, channel estimation in IRS-assisted systems is a challenging task because it is very difficult for passive elements to sense the channels.  Recently, there exist many works for passive estimation in IRS-aided systems with narrowband channels \cite{ON2}-\cite{sequence}. However, for wideband wireless communication systems, the problem will become quite complicated due to more channel coefficients to be estimated, which are brought by the multi-path delay spread. Up to now, only limited work focuses on access point (AP)-IRS-user cascade channel estimation in IRS-aided orthogonal frequency division multiplexing (OFDM) systems. Specifically, an ON/OFF-based protocol was firstly proposed in \cite{onoff}, in which only small portion of IRS elements are on at each time slot. However, this approach may suffer a large normalized-mean-square-error (NMSE) because the array gain cannot be fully exploited. In \cite{DFT}, the authors proposed a discrete Fourier transforms (DFT)-based reflecting pattern design to greatly reduce the estimation errors compared to the ON/OFF-based protocol. Moreover, the authors in \cite{fast} proposed a fast channel estimation scheme with reduced OFDM symbol duration to further reduce training overhead. Recently, machine learning (ML) based channel estimation strategies have also been investigated in \cite{ml2} to improve the channel estimation performance.

However, all the above works assume that the IRS has an ideal reflection model, i.e., constant amplitude, variable phase shifts, as well as the same response for the signals with different frequencies, which is impractical due to the real world hardware limitation. According to the analysis in our recent work \cite{w}, \cite{H}, the practical response (i.e., amplitude and phase shift) of IRS is tightly related to the frequency of the incident signal. When taking this practical model into consideration, many problems, especially the channel estimation strategy, should be reconsidered since existing work is not applicable anymore. Thus, it is very necessary to adopt a novel method to execute channel estimation based on the practical IRS model and  further reduce the NMSE of the estimated channels in IRS-assisted OFDM systems.

In this paper, we consider a simple point-to-point IRS-assisted OFDM system and  perform channel estimation based on our proposed practical IRS model. Next, aiming at minimizing the NMSE, we properly design an IRS time-varying reflection pattern  for the case of using low-resolution phase shifters by leveraging alternating optimization (AO) technique. Then, for high-resolution IRS cases, we propose a practical reflection pattern design scheme with lower computational complexity. Finally, simulation  results verify the necessity of taking the practical IRS model into consideration and the advantage of our proposed channel estimation methods.

\emph{Notations}: $(\cdot)^T$, $(\cdot)^H$, and $(\cdot)^{-1}$  denote  the transpose, the transpose-conjugate, and the inverse operations, respectively. $\| \mathbf{a} \|$ is the norm of a vector $\mathbf{a}$. $\odot$ denotes the Hadamard product. $\mathbb{E}\{\cdot\}$ represents statistical expectation. $\rm{Rank}(\cdot)$ and ${\rm{Tr}}\{\cdot\}$ denote the rank and the trace of matrix, respectively. $\mathbf{A}(i,j)$ denotes the element of the $i$-th row and the $j$-th column of  matrix $\mathbf{A}$. Finally, $\mathbf{a}(i)$ denotes the $i$-th element of vector $\mathbf{a}$.
\vspace{-0.08 cm}
\section{System Model}

\label{sec:system model}
\begin{figure}[!t]
\centering
\includegraphics[width = 3.1 in]{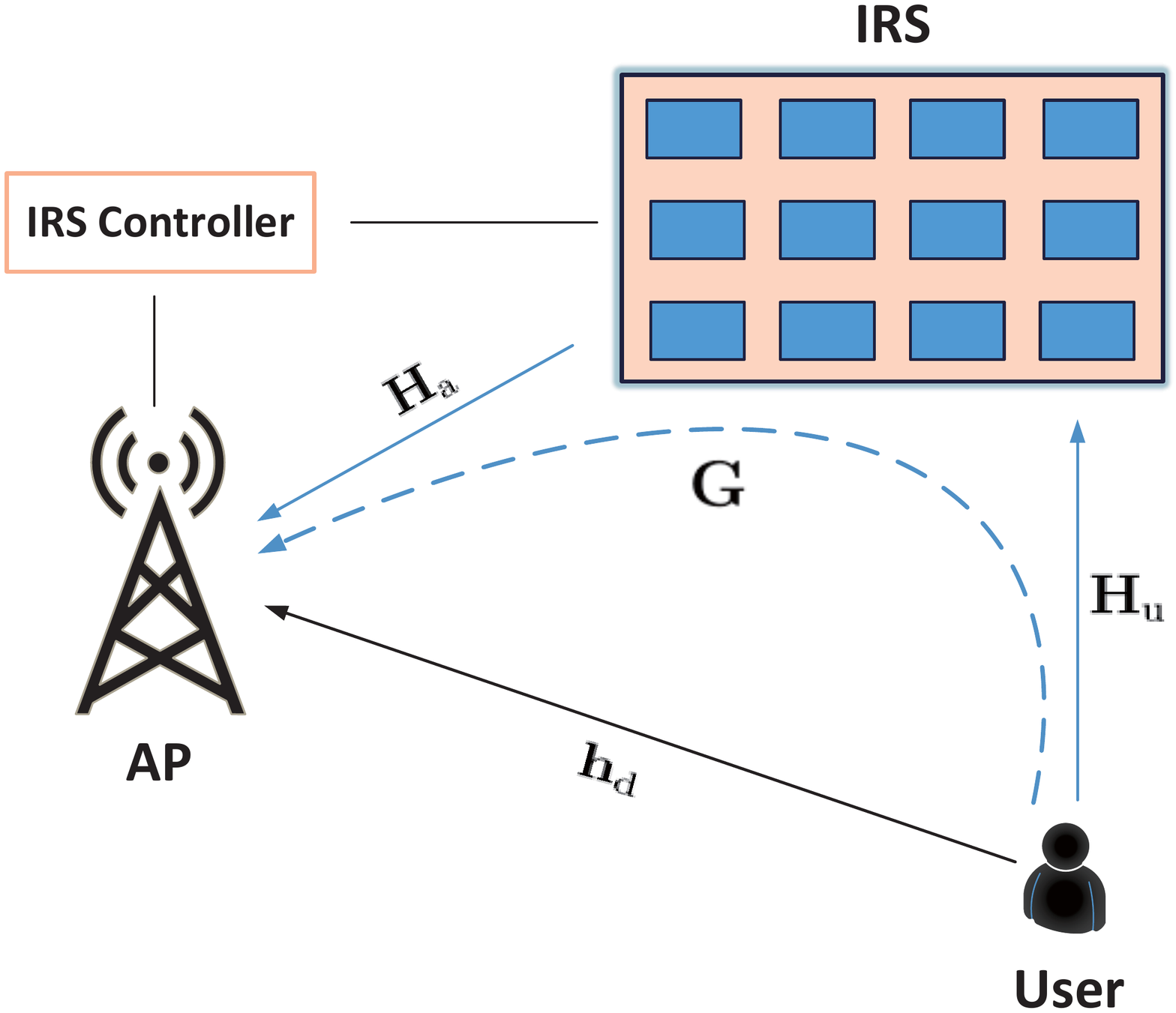}
\vspace{-0.2 cm}
\caption{An IRS-assisted uplink OFDM system.}
\label{fig:CEE}
\vspace{-0.7 cm}
\end{figure}

We consider an IRS-assisted uplink OFDM system with $N$ subcarriers, where both the AP and the user are equipped with single antenna as shown in Fig. \ref{fig:CEE}. By connecting with a smart controller, an IRS composed of $M$ elements can work collaboratively with the AP to intelligently enhance the propagation environment. Denote $\mathcal{N} = \{1,\cdots,N\}$ and $\mathcal{M} = \{1,\cdots,M\}$ as the sets of the indices of subcarriers and the elements of the IRS, respectively. Moreover, in this paper, we assume all the links to be quasi-static frequency-selective fading channels.

Let $\mathbf{H}_{\rm{a}}\in\mathbb{C}^{M{\times}N}$, $\mathbf{H}_{\rm{u}}\in\mathbb{C}^{M{\times}N}$, and $\mathbf{h}_{\rm{d}}\in\mathbb{C}^{N}$ denote the channel frequency response (CFR) of AP-IRS link, IRS-user link, as well as AP-user link, respectively.  $\mathbf{G}\triangleq\mathbf{H}_{\rm{u}} \odot \mathbf{H}_{\rm{a}}\in\mathbb{C}^{M\times N}$ denotes the CFR of user-IRS-AP cascade link (i.e., the dashed line in Fig. \ref{fig:CEE}). Let $\mathbf{x}=[x_{1},\cdots,x_{N}]^{T}{\in}\mathbb{C}^{N}$ denote the transmit OFDM symbol in frequency domain at the user side, and  $P_{\mathrm{t}}=\mathrm{Tr}\{\mathbb{E}\{\mathbf{x}^{H}\mathbf{x}\}\}$ denote the total transmit power. Then, the communication process can be described as follows: The transmit OFDM  symbol $\mathbf{x}$ is firstly transformed into time domain via an $N$-point inverse discrete fourier transform (IDFT), and then appended by a cyclic prefix (CP) of length $L_{\rm{cp}}$ to mitigate the inter-symbol-interference (ISI) in our considered wideband channels modeled by an $L$-tap finite-duration impulse response ($L_{\rm{cp}}\geq L$) \cite{DFT}. At the receiver side, by removing the CP and performing an $N$-point DFT, the typical formulation of equivalent frequency-domain received signal at the $n$-th subcarrier ${y}_{n}$, $\forall n \in \mathcal{N}$, can be given as:

\begin{equation}
\begin{aligned}
{y}_{n}&=(\mathbf{h}_{{\rm{u}},n}^{H}{\rm{diag}}(\boldsymbol{\phi}_{n})\mathbf{h}_{{\rm{a}},n}+{h}_{{\rm{d}},n})x_{n}+{v}_{n},\\
&=(\mathbf{g}_{n}^{H}\boldsymbol{\phi}_{n}+{h}_{{\rm{d}},n})x_{n}+{v}_{n},\\
&\mathop{=}\limits^{(a)}\mathbf{\widehat{g}}_{n}^{H}\boldsymbol{\psi}_{n}x_{n}+{v}_{n},
\label{eq:receive}
\end{aligned}
\end{equation}

\noindent where $h_{{\rm{d}},n}$ denotes the $n$-th element of $\mathbf{h}_{\rm{d}}$. $\mathbf{h}_{{\rm{u}},n}\in\mathbb{C}^{M}$, $\mathbf{h}_{{\rm{a}},n}\in\mathbb{C}^{M}$ and $\mathbf{g}_{n}\triangleq\mathbf{h}_{{\rm{u}},n}\odot \mathbf{h}_{{\rm{a}},n}\in \mathbb{C}^{M}$ are the $n$-th column of $\mathbf{H}_{\rm{u}}$, $\mathbf{H}_{\rm{a}}$ and $\mathbf{G}$, respectively. $v_{n}\sim\mathcal{CN}(0,\sigma^{2})$ is the additive Gaussian noise. Define $\boldsymbol{\phi}_{n}\triangleq[\beta_{n,1}e^{j\phi_{n,1}},\cdots,\beta_{n,M}e^{j\phi_{n,M}}]^{T}{\in}\mathbb{C}^{M}$ as the IRS reflection coefficients at the $n$-th subcarrier. Here, $\beta_{n,m}$ and $\phi_{n,m}$ denote the amplitude attenuation and phase shift generated by the $m$-th element corresponding to the $n$-th subcarrier, respectively.  And $(a)$ follows by defining $\mathbf{\widehat{g}}_{n}\triangleq[{h}_{{\rm{d}},n}^{H},\mathbf{g}_{n}^{H}]^{H}\in\mathbb{C}^{M+1}$ and $\boldsymbol{\psi}_{n}\triangleq\left[1,\boldsymbol{\phi}^{H}_{n}\right]^{H}\in\mathbb{C}^{M+1}$. Prior works on the channel estimation for IRS-assisted OFDM systems usually assume the IRS to have an ideal reflection model:  Reflection coefficients for different subcarriers are considered to be the same, i.e., $\beta_{1,m} = \cdots = \beta_{n,m}=1$, $\phi_{1,m} = \cdots = \phi_{N,m}$, $\forall m \in \mathcal{M}$. However, recent works \cite{w}, \cite{H} reveal that it is not appropriate and also impractical to make this simple assumption. Specifically, the electrical characteristics of an IRS element is illustrated by a parallel resonant circuit and its reflection coefficients are essentially determined by the equivalent impedance. The value of impedance not only depends on the circuit parameters (e.g., the equivalent capacitance), but also relates to the frequency of incident signal. This means that the same IRS element with certain circuit parameters generates different responses (i.e., reflection coefficients) to signals with different frequencies. Thus, for wideband signals, the amplitude and the phase shift of an IRS element must vary with different subcarriers. In \cite{w}, the authors have successfully analyzed the mathematical formulation between the reflection coefficients of IRS and the frequency of the incident signal. Specifically, if we know the phase shift of an element at the carrier frequency $f_{\mathrm{c}}$ of the total bandwidth $B$, then the corresponding reflection coefficients of the same element at different subcarriers can be derived subsequently, which is given by \cite{w}:
\begin{subequations}
\begin{align}
&A(\phi_{{\rm{c}}},f_{n})=-\frac{\alpha_{4} \phi_{{\rm{c}}}+\alpha_{5}}{\left(0.05\left(f_{n} / 10^{9}-\mathcal{D}_{1}\left(\phi_{{\rm{c}}}\right)\right) \right)^{2}+4}+1 \label{eq:w1}, \forall n, \\
&W(\phi_{{\rm{c}}},f_{n}) = -2 \tan ^{-1}[\mathcal{D}_{2}\left(\phi_{{\rm{c}}}\right)\left(f_{n} / 10^{9}-\mathcal{D}_{1}\left( \phi_{\mathrm{c}}\right)\right)], \forall n,\\
&\mathcal{D}_{1}(\phi_{\mathrm{c}}) = \alpha_{1}\tan(\phi_{\mathrm{c}}/3)+\alpha_{2}\sin(\phi_{\mathrm{c}})+\alpha_{6}, \forall n ,  \label{eq:w2}\\
&\mathcal{D}_{2}(\phi_{\mathrm{c}}) = \alpha_{3}\phi_{\mathrm{c}}+\alpha_{7}, \forall n,\label{eq:wfinal}
\end{align}
\end{subequations}
where  $\phi_{\mathrm{c}}$ denotes the phase shift corresponding to the carrier frequency $f_{\mathrm{c}}$. $f_{n}$ denotes the central frequency at the $n$-th subcarrier with $f_{n} \triangleq f_{\mathrm{c}}+\left(i-\frac{N+1}{2}\right) \frac{B}{N}, \forall n \in \mathcal{N}$. $A(\phi_{{\rm{c}}},f_{n})$ and $W(\phi_{{\rm{c}}},f_{n})$ denote the amplitude and phase shift as a function of $\phi_{\mathrm{c}}$ and $f_{n}$ for the $n$-th subcarrier, respectively.  ${\{\alpha_{i}\}}_{i=1}^{7}$ are the parameters related to specific circuit implementation. In practice, due to practical hardware implementation, the phase shift of each reflecting element always has only a finite number of discrete values \cite{003}. Thus, in this paper we assume that each reflecting element has finite-resolution phases controlled by $b$ bits, which yields a finite set: \begin{equation}\mathcal{F}_{b}=\{0,\Delta\omega,\cdots,(2^{b}-1)\Delta\omega\},\label{eq:F}\end{equation}
where $\Delta\omega={2\pi}/{2^{b}}$. Based on the above discussion, we further expand the signal model in the time domain, which will be utilized for the following channel estimation. Specifically, we define $K$ time slots at each subcarrier and let $x_{n,k}$ denote the transmit symbol corresponding to the $n$-th subcarrier at the $k$-th time slot, $\forall n \in \mathcal{N}, k = 1,\cdots,K$. Then the received signal $\widehat{y}_{n,k}$ in our considered practical model during the $k$-th time slot at the $n$-th subcarrier can be given as:
\begin{equation}
\widehat{y}_{n,k}= \mathbf{\widehat{g}}_{n}^{H}\boldsymbol{\psi}_{n,k}x_{n,k}+v_{n,k}, \forall n, \forall k,\label{eq:receive2}
\end{equation}
where $\boldsymbol{\psi}_{n,k}$ denotes the time-expanded reflection vector corresponding to the $n$-th subcarrier and the $k$-th time slot, which is defined as: $\boldsymbol{\psi}_{n,k}\triangleq[1,\boldsymbol{\phi}_{n,k}^{H}]^{H}, \boldsymbol{\phi}_{n,k}\triangleq[\beta_{n,1,k}e^{j\phi_{n,1,k}},\cdots,\beta_{n,M,k}e^{j\phi_{n,M,k}}]^{T}$. $\beta_{n,m,k}$ and $\phi_{n,m,k}$ denote the amplitude and phase shift of the $m$-th element at the $n$-th subcarrier during the $k$-th time slot, respectively, and they follow the relationship given in (\ref{eq:w1})-({\ref{eq:wfinal}), i.e., $\beta_{n,m,k}=A(\phi_{\mathrm{c},m,k},f_{n})$, $\phi_{n,m,k}=W(\phi_{\mathrm{c},m,k},f_{n})$, $\forall n \in \mathcal{N}, \forall m \in \mathcal{M}, k = 1,\cdots,K$. $\phi_{\mathrm{c},m,k}$ denotes the phase shift of the $m$-th element corresponding to $f_{\mathrm{c}}$ at the $k$-th time slot. $v_{n,k}$ is the  noise component at the $n$-th subcarrier during the $k$-th time slot.

In this paper, we aim to estimate the CFR of the combination of cascade link and direct link at all the subcarriers, i.e., $\mathbf{\widehat{g}}_{n}$, $\forall n \in \mathcal{N}$, at the AP. In the uplink training phase, the user consecutively sends $K$ pilot symbols $x_{n,k}, k = 1,\cdots,K$, each of which is associated with the pre-designed IRS reflection pattern (i.e., $\boldsymbol{\psi}_{n,k}$). Then the AP processes $\widehat{y}_{n,k}$ based on $x_{n,k}$ and $\boldsymbol{\psi}_{n,k}$, $\forall n \in \mathcal{N}$, $k=1,\cdots K$, to perform the channel estimation.  Since the accuracy of the estimated channel is tightly associated with the response of the IRS,  the time-varying reflection pattern (i.e., $\boldsymbol{\psi}_{n,k}, \forall n \in \mathcal{N}, k=1,\cdots,K$) of the IRS needs to be properly designed to reduce the estimation errors. In the following two sections, we will describe the complete procedure of the channel estimation and provide the solutions of the reflection pattern design in details.

\vspace{-0.05 cm}
\section{Proposed Channel Estimation Method and NMSE Analysis}

In this section, we propose a channel estimation method based on the practical IRS reflection model and provide analysis about the NMSE of the estimated channels.
\vspace{-0.2 cm}
\subsection{ Proposed Channel Estimation Method}
Before deriving the estimate of cascade link's CFR, we firstly need to perform the  estimation of $\mathbf{\widehat{g}}_{n}\boldsymbol{\psi}_{n}$ during each training time slot. By right multiplying ${x}_{n,k}^{-1}$ on (\ref{eq:receive2}), the  estimate of $\mathbf{\widehat{g}}_{n}\boldsymbol{\psi}_{n,k}$, denoted by $\widehat{z}_{n,k}$, can be formulated as:
\begin{equation}
\widehat{z}_{n,k}=\mathbf{\widehat{g}}_{n}^{H}\boldsymbol{\psi}_{n,k}+{v}_{n,k}{x}_{n,k}^{-1},\forall n, \forall k.
\label{eq:LS1}
\end{equation}
Next, we seek to extract the estimate of $\mathbf{\widehat{g}}_{n}$ from $\widehat{z}_{n,k}$, $\forall n \in \mathcal{N}$, $k=1,\cdots,K$. Based on the discussion in \cite{ls}, we know that the number of transmitted pilot symbols needs to be no less than $M+1$, so that the LS algorithm can be performed successfully. Therefore, the transmission protocol can be expressed as follows: \textit{i)} The user consecutively sends $K=M+1$ OFDM symbols within one channel coherence time; \textit{ii)} next, the AP receives the signal reflected by the IRS; \textit{iii)} finally, the AP obtains $\widehat{z}_{n,k}$, $\forall n \in \mathcal{N}$, for each time slot based on (\ref{eq:LS1}) and aggregate them together. Since the reflection pattern $\boldsymbol{\psi}_{n,k}, \forall n \in \mathcal{N}, k = 1,\cdots K,$ are pre-defined, the estimation of $\mathbf{\widehat{g}}_{n}$ can be obtained at the AP directly based on the above protocol. In particular, we define $\mathbf{\widehat{z}}_{n}\triangleq[\widehat{z}_{n,1},\cdots,\widehat{z}_{n,K}]^{H}, \forall n \in \mathcal{N},$ as the aggregated $\widehat{z}_{n,k}$, which can be formulated as:
\begin{equation}
\mathbf{\widehat{z}}_{n}=\boldsymbol{\Psi}_{n}^{H}\mathbf{\widehat{g}}_{n}+\mathbf{\widehat{v}}_{n}, \forall n,
\label{eq:LS2}
\end{equation}
where
\begin{equation}
\boldsymbol{\Psi}_{n}=[\boldsymbol{\psi}_{n,1},\cdots,\boldsymbol{\psi}_{n,K}], \forall n,
\label{eq:rank2}
\end{equation}
\begin{equation}
\begin{aligned}
\widehat{\mathbf{v}}_{n}&=[v_{n,1}{x}_{n,1}^{-1},\cdots,{v}_{n,k}x_{n,K}^{-1}]^{H}\\
&\mathop{=}\limits^{(b)}({\rm{diag}}(\mathbf{\widehat{x}}_{n}^{-1}))^{H}\mathbf{\widetilde{v}}_{n}, \forall n,
\end{aligned}
\end{equation}
and ($b$) holds by defining $\mathbf{\widetilde{v}}_{n}\triangleq[v_{n,1},\cdots,v_{n,K}]^{H}$, $\mathbf{\widehat{x}}_{n}\triangleq[x_{n,1},\cdots,x_{n,K}]^{H}$.
If $\boldsymbol{\Psi}_{n}$ is  full-rank, the LS estimation of $\mathbf{\widehat{g}}_{n}$ is performed by multiplying $(\boldsymbol{\Psi}_{n}^{H})^{-1}$ on (\ref{eq:LS2}), which yields the estimated channel $\widetilde{\mathbf{g}}_{n}$ as:
\begin{equation}
\begin{aligned}
\widetilde{\mathbf{g}}_{n}&=(\boldsymbol{\Psi}_{n}^{H})^{-1}\mathbf{\widehat{z}}_{n}
=\mathbf{\widehat{g}}_{n}+(\boldsymbol{\Psi}_{n}^{H})^{-1}\mathbf{\widehat{v}}_{n}, \forall n.
\end{aligned}
\label{eq:LSfinal}
\end{equation}
Now the protocol to estimate the CFR of cascade link and direct link for all subcarriers is established. In the following subsection, we will provide theoretical derivation of the NMSE of estimated channels.
\vspace{-0.3cm}
\subsection{NMSE Analysis}

In this subsection, we derive the NMSE of estimated channels, the formulation of which can be given as:
\begin{equation}
{\rm{NMSE}}= \frac{1}{N} \mathbb{E}\left\{\frac{\sum\limits_{n=1}^{N}\|\widetilde{\mathbf{g}}_{n}-\mathbf{\widehat{g}}_{n}\|^{2}}{\sum\limits_{n=1}^{N}\|\mathbf{\widehat{g}}_{n}\|^{2}}\right\}.
\label{eq:NMSE00}
\end{equation}
Substituting equations (\ref{eq:LS1})-(\ref{eq:LSfinal}) into (\ref{eq:NMSE00}), (\ref{eq:NMSE00}) can be transformed as:
\begin{equation}
\begin{aligned}
{\rm{NMSE}}&=\frac{1}{N} \frac{\sum\limits_{n=1}^{N}\mathbb{E}\{\|\widetilde{\mathbf{g}}_{n}-\mathbf{\widehat{g}}_{n}\|^{2}\}}{\sum\limits_{n=1}^{N}\mathbb{E}\{\|\mathbf{\widehat{g}}_{n}\|^{2}\}}\\
&=\frac{1}{N}\frac{\sum\limits_{n=1}^{N}{\rm{Tr}}\left\{\mathbb{E}\{(\boldsymbol{\Psi}_{n}^{-1})^{H}\mathbf{\widehat{v}}_{n}\mathbf{\widehat{v}}_{n}^{H}\boldsymbol{\Psi}_{n}^{-1}\}\right\}}{{\sum\limits_{n=1}^{N}\mathbb{E}\{\|\mathbf{\widehat{g}}_{n}\|^{2}\}}}.
\end{aligned}
\label{eq:NMSE01}
\end{equation}
The numerator of (\ref{eq:NMSE01}) can be further rewritten as:
\begin{equation}
\begin{aligned}
{\rm{Tr}}\left\{\mathbb{E}\{(\boldsymbol{\Psi}_{n}^{-1})^{H}\mathbf{\widehat{v}}_{n}\mathbf{\widehat{v}}_{n}^{H}\boldsymbol{\Psi}_{n}^{-1}\}\right\}&={\rm{Tr}}\{({\boldsymbol{\Psi}_{n}^{-1}})^{H}\mathbb{E}\{({\rm{diag}}(\mathbf{\widehat{x}}^{-1}_{n}))^{H}\\
&~\times\mathbf{\widetilde{v}}_{n}(\mathbf{\widetilde{v}}_{n})^{H}{\rm{diag}}(\mathbf{\widehat{x}}^{-1}_{n})\}{\boldsymbol{\Psi}_{n}^{-1}}\}\\
&=\frac{\sigma^{2}}{P_{n}}{\rm{Tr}}\left\{({\boldsymbol{\Psi}_{n}^{-1}})^{H}{\boldsymbol{\Psi}_{n}^{-1}}\right\},
\end{aligned}
\label{eq:TR1}
\end{equation}
where $P_{n}$ denotes the average transmit power for the $n$-th subcarrier. Here, we assume the transmit power is equally allocated in each subcarrier, i.e., $P_{n}=\frac{P_{\mathrm{t}}}{N}$. Thus, the NMSE (\ref{eq:NMSE01}) can be reformulated as:
\begin{equation}
{\rm{NMSE}}=\frac{\sigma^{2}}{P_{\mathrm{t}}}\frac{\sum\limits_{n=1}^{N}{\rm{Tr}}\{({\boldsymbol{\Psi}_{n}^{-1}})^{H}{\boldsymbol{\Psi}_{n}^{-1}}\}}{{\sum\limits_{n=1}^{N}\mathbb{E}\{\|\mathbf{\widehat{g}}_{n}\|^{2}\}}}.
\label{eq:MSE}
\end{equation}

From (\ref{eq:MSE}), we can conclude that the value of the NMSE is heavily determined by the reflection pattern $\boldsymbol{\Psi}_{n}$. If the reflection pattern ${\boldsymbol{\Psi}_{n}}$ is not properly constructed, the NMSE will become quite high. Therefore, how to design a proper IRS time-varying reflection pattern is a crucial issue for channel estimation, which will be discussed in the next section.

\section{Reflection Pattern Design}

In this section we address the problem of the IRS time-varying reflection pattern design to minimize the NMSE. Although some recent works \cite{onoff}-\cite{fast} have investigated proper pattern design schemes, they mainly focus on channel estimation strategies based on ideal IRSs. When the practical IRS-assisted wireless communication system is considered, these methods cannot suit well anymore. From (\ref{eq:MSE}), we know that the value of NMSE is related to  ${\boldsymbol{\Psi}_{n}}$ corresponding to different subcarriers. However, $\boldsymbol{\Psi}_{n}, \forall n \in \mathcal{N},$ cannot be designed individually since they are interrelated. Specifically, during each training time slot, the response of practical IRS is tuned according to one frequency (i.e., selecting an appropriate capacitance), which yields different responses to signals with different frequencies. Here, we consider the reflection pattern design focusing on the carrier frequency of the total bandwidth (i.e., $f_{\mathrm{c}}$). Therefore, our objective is formulated as properly designing the reflection pattern at $f_{\mathrm{c}}$ to minimize the NMSE derived in the previous section. Let $\boldsymbol{\Phi}_{\mathrm{c}}$ denote the time-varying reflection pattern matrix corresponding to $f_{\mathrm{c}}$, $\boldsymbol{\Phi}_{\mathrm{c}}\triangleq[\boldsymbol{\theta}_{1},\cdots,\boldsymbol{\theta}_{K}]$, $\boldsymbol{\theta}_{k}\triangleq[1,\phi_{\mathrm{c},1,k},\cdots,\phi_{\mathrm{c},M,k}]^{T}$,  $k=1,\cdots,K$. Then, the reflection pattern design problem is given by:
\begin{subequations}
\begin{align}
\label{eq:15a}
&\mathop{\rm{min}}\limits_{{\boldsymbol{\Phi}_{\mathrm{c}}}}\sum\limits_{n=1}^{N}{\rm{Tr}}\{({\boldsymbol{\Psi}_{n}^{-1}})^{H}{\boldsymbol{\Psi}_{n}^{-1}}\}\\
&{\rm{s.t.}}~\text{Rank}({\boldsymbol{\Psi}_{n}}) = M+1, \forall n,\label{eq:zhi}\\
&~~~~\boldsymbol{\Psi}_{n}(1,:)=\mathbf{1}_{1 \times (M+1)}, \forall n,\label{eq:one2}\\
&~~~~\beta_{n,m,k}=A(\phi_{\mathrm{c},m,k},f_{n}), \forall n, \forall m, \forall k, \label{eq:14d}\\
&~~~~\phi_{n,m,k}=W(\phi_{\mathrm{c},m,k},f_{n}), \forall n, \forall m, \forall k, \label{eq:14e}\\
&~~~~\phi_{\mathrm{c},m,k} \in \mathcal{F}_{b}, \forall m, \forall k, \label{eq:phic}
\end{align}
\end{subequations}
where (\ref{eq:zhi}) is a restriction to assure the existence of $\boldsymbol{\Psi}_{n}^{-1}$, and (\ref{eq:one2}) is derived based on the definition of  $\boldsymbol{\psi}_{n}$. According to (\ref{eq:14d}) and (\ref{eq:14e}), $\boldsymbol{\Psi}_{n}$, $\forall n \in \mathcal{N}$, can be correspondingly calculated when $\boldsymbol{\Phi}_{\mathrm{c}}$ is given. Obviously, this is a non-convex and complicated problem, which cannot be solved directly. To effectively and efficiently cope with this difficulty, we firstly obtain the solution of $\boldsymbol{\Phi}_{\mathrm{c}}$ by leveraging AO \cite{ao} technique for the case of low-resolution IRS (i.e., $b$ = 1, 2). Then, we extend our analysis for high-resolution IRS case (i.e., $b \geq 3$) and modify the practical reflection pattern design scheme to efficiently calculate $\boldsymbol{\Phi}_{\mathrm{c}}$.
\vspace{-0.25 cm}
\subsection{Low-Resolution IRS Cases}

In this subsection, we begin with using AO algorithm to find the optimal ${\boldsymbol{\Phi}_{\mathrm{c}}}$  to construct the desired reflection pattern for low-resolution IRS cases. Given an initial value of $\boldsymbol{\Phi}_{\mathrm{c}}$, we aim to successively update each element of the reflection pattern matrix. Specifically, for the design of $\boldsymbol{\Phi}_{\mathrm{c}}(i,j), i = 2,\cdots,M+1, j = 1,\cdots,M+1$, we attempt to conditionally minimize the NMSE with fixed other elements of $\boldsymbol{\Phi}_{\mathrm{c}}$, i.e.,
\begin{equation}
\boldsymbol{\Phi}_{\mathrm{c}}(i,j) = \arg \min_{\boldsymbol{\Phi}_{\mathrm{c}}(i,j) \in \mathcal{F}_{b}} \sum_{n=1}^N {\rm{Tr}}\{({\boldsymbol{\Psi}_{n}^{-1}})^{H}{\boldsymbol{\Psi}_{n}^{-1}}\} , \forall i, \forall j.\label{eq:16a}
\end{equation}
Thanks to the employment of low-resolution phase shifters for practical IRS realization, we can perform a low-complexity one-dimensional exhaustive search over the set $\mathcal{F}_{b}$. The above process will be repeatedly executed until the number of iterations equals to $S_{max}$. Here, $S_{max}$ is set properly to guarantee the convergence of the total procedure, and the value of it will be discussed in Section V. It is worth noting that the initial value of the reflection pattern will, to a large extent, influence the performance of this kind of AO algorithm. Thus, we also need to set an appropriate initial $\boldsymbol{\Phi}_{\mathrm{c}}^{(0)}$. In this paper, we utilize the DFT-hadamard matrix scheme for discrete phase shifts in \cite{baseline2} as an initial value of $\boldsymbol{\Phi}_{\mathrm{c}}$. The above procedure is summarized in Algorithm 1.
\begin{algorithm}[!t]
\small
\caption{AO-based Low-Resolution Reflection Pattern Design}
\label{alg:1}
    \begin{algorithmic}[1]
    \REQUIRE $M$, $N$, $b$.
    \ENSURE $\boldsymbol{\Phi}_{\mathrm{c}}$.
        \STATE {Initialize $\boldsymbol{\Phi}_{\mathrm{c}}^{(0)}$ according to \cite{baseline2}.}
            \FOR {$iter=1$ : $S_{max}$}
                \FOR {$i$ = 2 to $M+1$ }
                   \FOR {$j$ = 1 to $M+1$}
                      \STATE{Update $\boldsymbol{\Phi}_{\mathrm{c}}(i,j)$ by solving problem (\ref{eq:16a}).}
                 \ENDFOR
                \ENDFOR
            \ENDFOR
            \STATE{$\boldsymbol{\Phi}_{\mathrm{c}}$ = $\boldsymbol{\Phi}_{\mathrm{c}}^{(iter)}$.}
    \end{algorithmic}
\end{algorithm}

Although the above AO technique is superior compared to exhaustive search, at least $2^{b}M(M+1)$ searching operations are needed to obtain the solution of $\boldsymbol{\Phi}_{\mathrm{c}}$, which will cause high computational complexity if $b$ is larger (i.e., $b \geq 3$), Thus, in the following subsection, we attempt to further modify the reflection pattern design based on our proposed AO-based algorithm in order to better suit for high-resolution IRS cases.
\vspace{-0.3 cm}
\subsection{High-Resolution IRS Cases}

In this subsection, we propose to further modify Algorithm 1 based on numerous numerical experiments to effectively reduce the complexity for the  practical high-resolution reflection pattern design scheme. Specifically, we attempt to explore the characteristic of the reflection pattern matrix obtained by Algorithm 1 and find  the insightful relationship between $(b-1)$-bit reflection matrix $\boldsymbol{\Phi}_{\mathrm{c}}^{b-1}$ and the $b$-bit one $\boldsymbol{\Phi}_{\mathrm{c}}^{b}$, which can be expressed as: If we obtain a certain angle $\mathcal{F}_{b}[q]$, which denotes the $q$-th element of the set $\mathcal{F}_{b}$, as the $(i,j)$-th element of $\boldsymbol{\Phi}_{\mathrm{c}}^{b-1}$ based on Algorithm 1, i.e., $\boldsymbol{\Phi}_{\mathrm{c}}^{b-1}(i,j) = \mathcal{F}_{b}[q]$, then the value of $\boldsymbol{\Phi}_{\mathrm{c}}^{b}(i,j)$ is always within a set $\mathcal{S}_{q,b} \triangleq \{\mathcal{F}_{b}[q-1],\mathcal{F}_{b}[q],\mathcal{F}_{b}[q+1]\}$, which includes the angle $\mathcal{F}_{b}[q]$ and its neighbors.  Motivated by this finding, we can efficiently search the optimal $\boldsymbol{\Phi}_{\mathrm{c}}^{b}$ based on $\boldsymbol{\Phi}_{\mathrm{c}}^{b-1}$ and the corresponding set $\mathcal{S}_{q,b}$, i.e.,
\begin{equation}
\boldsymbol{\Phi}_{\mathrm{c}}(i,j) = \arg \min_{\boldsymbol{\Phi}_{\mathrm{c}}(i,j) \in \mathcal{S}_{q,b}} \sum_{n=1}^N {\rm{Tr}}\{({\boldsymbol{\Psi}_{n}^{-1}})^{H}{\boldsymbol{\Psi}_{n}^{-1}}\} , \forall i, \forall j. \label{eq:17a}
\end{equation}
By repeating this step over all elements, the reflection pattern matrix can be obtained. The procedure is fully described in Algorithm 2.  In this way, we only need $2bM(M+1)$ searching operations to construct $\boldsymbol{\Phi}_{\mathrm{c}}$ instead of $2^{b}$, which can significantly reduce the complexity as compared to Algorithm 1.

\begin{algorithm}[!t]
\small
\caption{Practical High-Resolution Reflection Pattern Design}
\label{alg:2}
    \begin{algorithmic}[1]
    \REQUIRE $M$, $N$, $b$.
    \ENSURE $\boldsymbol{\Phi}_{\mathrm{c}}$.
        \STATE {Initialize $\boldsymbol{\Phi}_{\mathrm{c}} = \boldsymbol{\Phi}_{\mathrm{c}}^{2}$.}
            \FOR {$b_{1}=3$ : $b$}
                \STATE{Update $\mathcal{F}_{b_{1}}$ by (\ref{eq:F}).}
                   \FOR {$i=2$ : $(M+1)$}
                     \FOR {$j=1$ : $(M+1)$}
                        \STATE{Find $q$ with $\mathcal{F}_{b_{1}}[q] = \mathbf{\Phi}_{\mathrm{c}}(i,j)$.}
                        \STATE{Update $\mathcal{S}_{q,b_{1}}$.}
                        \STATE{Obtain $\mathbf{\Phi}_{\mathrm{c}}(i,j)$ by solving problem (\ref{eq:17a}).}
                   \ENDFOR
                \ENDFOR
            \ENDFOR
    \end{algorithmic}
\end{algorithm}

\section{Simulation Results}

In this section, some simulation results are provided to validate the effectiveness of our proposed channel estimation methods and reflection pattern design schemes. We consider an IRS-aided single-user OFDM system with an IRS of $M=36$ elements. Both the distances of AP-IRS link and AP-user link are set as 50 meters, while the distance between IRS and user is 2 meters. Moreover, we assume that all the links follow Rayleigh fading and the value of the reference channel power gain at a distance of 1 m is set as -30dB. The pass loss exponents of AP-IRS link, IRS-user link, and AP-user link are set as 2.2, 2.4, and 3.5, respectively. The noise power $\sigma^{2}$ is set as -80dBm. The transmit OFDM symbol at the $n$-th subcarrier during the $k$-th time slot satisfies $x_{n,k} = \sqrt{{P_{\mathrm{t}}}/{N}}e^{jw_{n,k}}$, where $w_{n,k}$ is
the random phase uniformly distributed within the range of [0,2$\pi$). The specific values of ${\{\alpha_{i}\}}_{i=1}^{7}$ are set according to \cite{w}. Other parameters are set as follows: $f_{\mathrm{c}}=2.4$GHz, $N = 64$, $B=0.2$GHz, $L = 8$, and $L_{cp} = 16$.

We firstly examine the necessity of taking the practical response of IRS into consideration for  channel estimation. In Fig. \ref{fig:frequency2}, we plot the NMSE versus transmit power under our proposed practical channel estimation methods by using the existing reflecting pattern design for discrete phase shifts (i.e., DFT-hadamard-matrix design proposed in \cite{baseline2}). In addition, we perform channel estimation by using the prior channel estimation method, i.e., the signals for different frequencies are considered to share the same reflection pattern. As the transmit power becomes larger, it is observed that the NMSE decreases drastically when executing our proposed practical channel estimation method. By contrast, the NMSE based on the prior channel estimation methods achieves a worse performance and the value of its NMSE does not decrease effectively with  increasing transmit power. The gap between these two schemes becomes even larger with the growth of transmit power, which demonstrates the necessity of considering practical IRS model for channel estimation.
\begin{figure}[!t]
\centering
\includegraphics[width = 3.45 in]{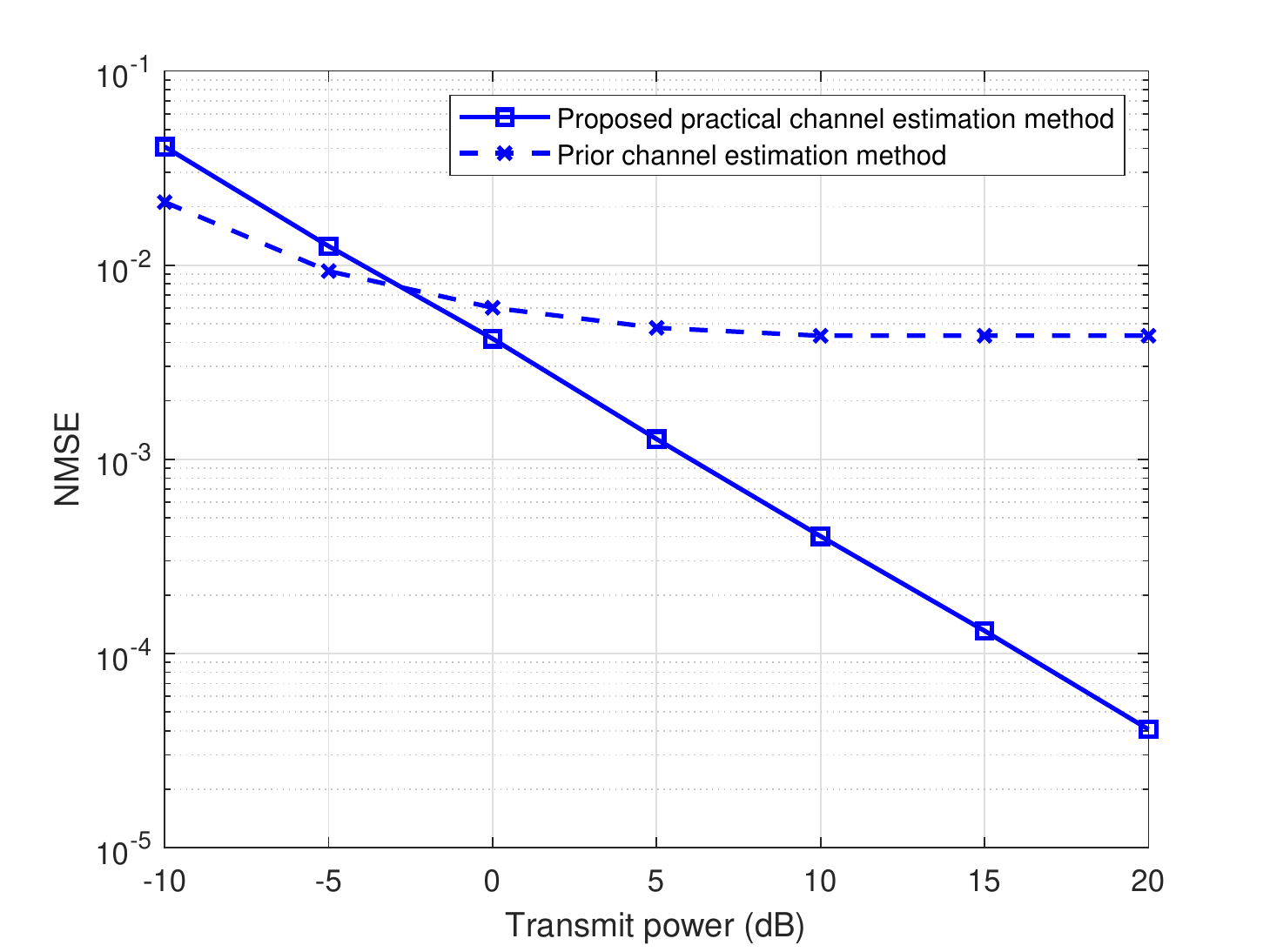}
\vspace{-0.1 cm}
\caption{NMSE versus transmit power under existing reflection pattern design schemes ($b=5$).}
\label{fig:frequency2}
\vspace{-0.3 cm}
\end{figure}

Next, we evaluate the covergence of our proposed Algorithm 1 by plotting the value of (\ref{eq:15a}) versus the number of iterations for the case of employing 1-bit resolution phase shifters. In the top sub-figure of Fig. \ref{fig:A2}, it can be observed that Algorithm 1 can converge within limited iterations, (i.e., within 2 iterations). Thus, in this paper, we set $S_{max}=2$ directly to improve the calculating efficiency. Then, in the bottom sub-figure of Fig. 3, we examine the effectiveness of our proposed Algorithm 2 and plot the value of objective (\ref{eq:15a}) as a function of resolution $b$ for each IRS element. From this sub-figure we can see that $b = 5$ is a sufficient resolution since the decrease of NMSE is marginal when $b \geq 5$.

Finally, we show the superior performance of our proposed reflection pattern design schemes. In Fig. \ref{fig:NMSE2}, we plot the NMSE versus transmit power for the scenarios with different resolution bits, under our proposed AO-based low-resolution reflection pattern design (i.e., Algorithm 1) and practical high-resolution reflection pattern design (i.e., Algorithm 2). For comparison, we also add DFT-hadamard-matrix based design proposed in \cite{baseline2} as the benchmark. It can be observed in Fig. \ref{fig:NMSE2} that the proposed schemes always achieve much better performance for all transmit power ranges. Compared with the DFT-hadamard-matrix based method, our proposed reflection pattern design schemes can dramatically reduce the NMSE, e.g., decrease by 46 $\%$ and 37 $\%$ for low-resolution IRS cases ($b=1$) and  high-resolution IRS cases ($b \geq 3$), respectively. Moreover, to further demonstrate the feasibility of the proposed practical high-resolution reflection pattern design, we also compare the NMSE by using proposed Algorithm 1 to obtain the solution of reflecting pattern when $b=3$. It can be observed that our proposed practical reflection pattern design Algorithm 2 can achieve almost the same performance as Algorithm 1, but can significantly reduce the computational complexity owing to the narrower searching range, especially for the systems with high resolution IRS, which, again, illustrate the effectiveness and efficiency of our proposed design.

\begin{figure}[!t]
\centering
\includegraphics[width = 3.45 in]{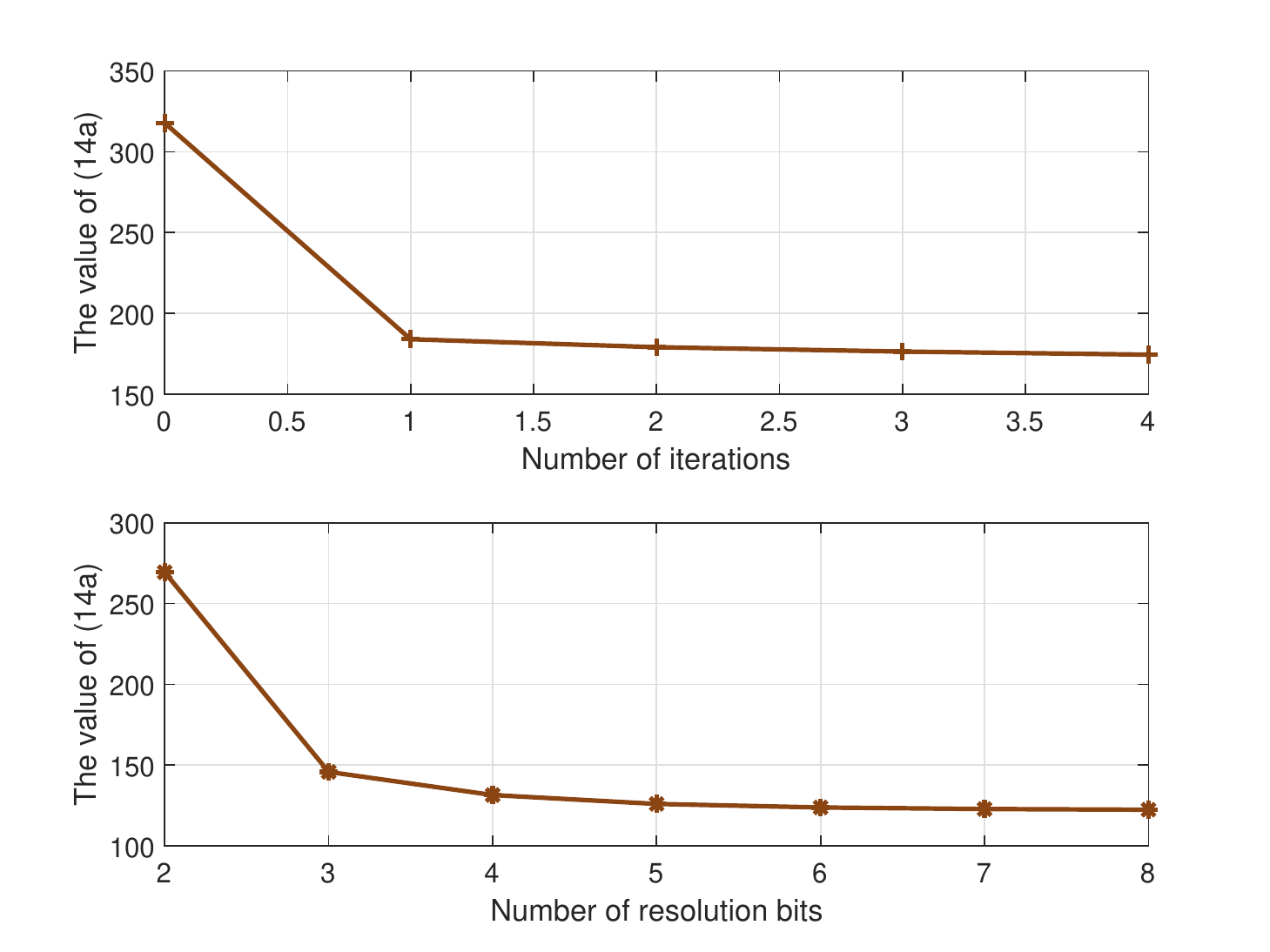}
\vspace{-0.1 cm}
\caption{Top: The value of (14a) versus the number of iterations based on Algorithm 1, $b=1$. Bottom: The value of (14a) versus the resolution $b$ based on Algorithm 2.}
\label{fig:A2}
\vspace{-0.3 cm}
\end{figure}

\begin{figure}[!t]
\centering
\includegraphics[width = 3.45 in]{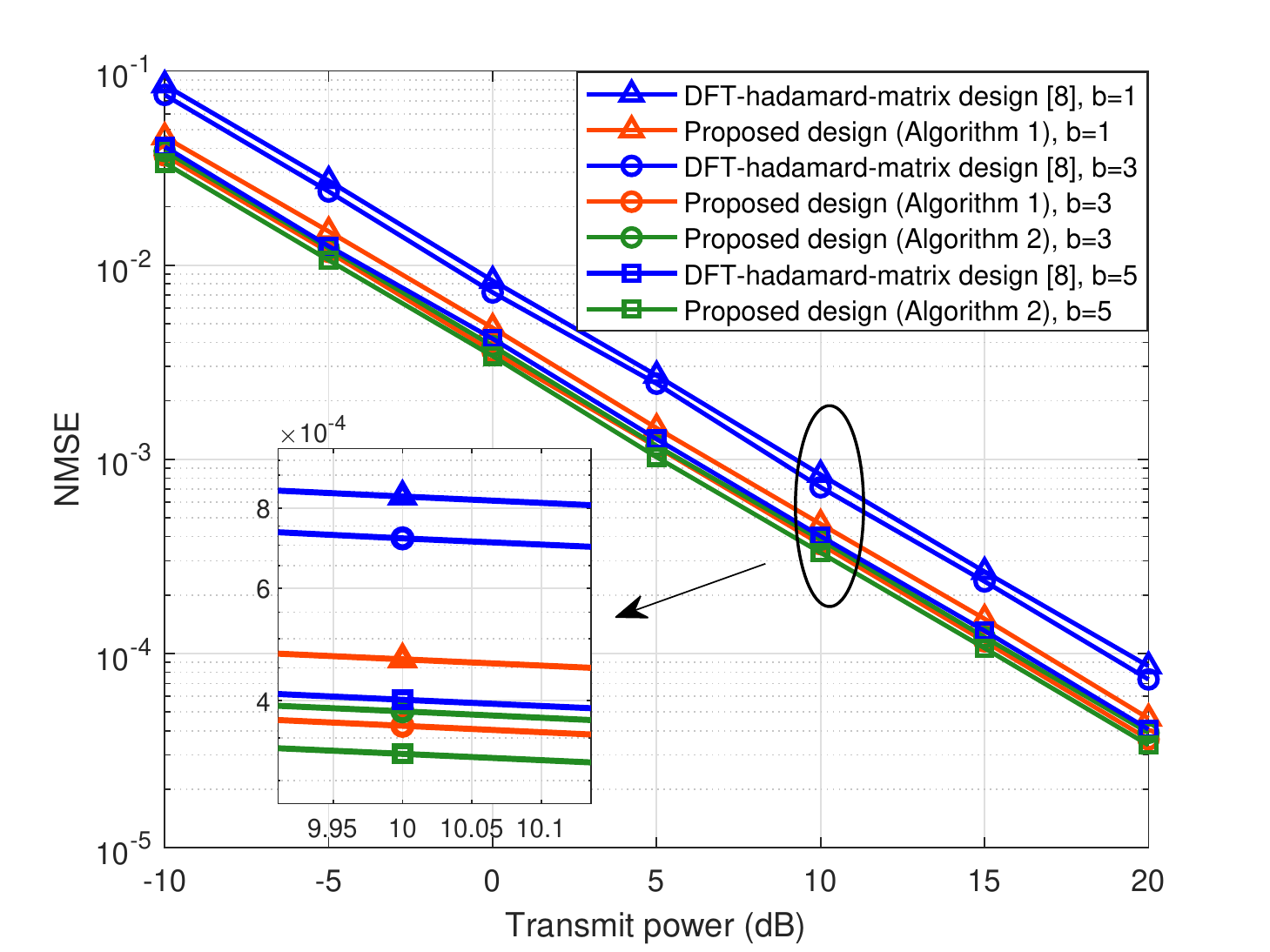}
\vspace{-0.1 cm}
\caption{NMSE versus transmit power for low-resolution IRS cases and high-resolution IRS cases.}
\label{fig:NMSE2}
\vspace{-0.3 cm}
\end{figure}

\section{Conclusions}

In this paper, we first proposed a novel channel estimation method for IRS-assisted wideband OFDM systems by considering IRS with a practical reflection model. Then, we designed a proper IRS time-varying reflecting pattern based on AO algorithm to minimize the NMSE for low-resolution IRS cases. To further improve the computational efficiency in designing reflecting pattern for the case of using high resolution phase shifters, another practical high-resolution reflection pattern design scheme was proposed. Simulation results demonstrated the effectiveness of our proposed methods.

\enlargethispage{-6.5cm}
\end{document}